\def\expec#1{\langle#1\rangle}

\def\nothing{\noindent\centerline{\,}}

\def\etal{{\frenchspacing et al.}}
\def\ie{{\frenchspacing i.e.}}

\def\beq#1{\begin{equation}\label{#1}}
\def\eeq{\end{equation}}
\def\beqa#1{\begin{eqnarray}\label{#1}}
\def\eeqa{\end{eqnarray}}

\def\spose#1{\hbox to 0pt{#1\hss}}
\def\simlt{\mathrel{\spose{\lower 3pt\hbox{$\mathchar"218$}}
     \raise 2.0pt\hbox{$\mathchar"13C$}}}
\def\simgt{\mathrel{\spose{\lower 3pt\hbox{$\mathchar"218$}}
     \raise 2.0pt\hbox{$\mathchar"13E$}}}
\def\simpropto{\mathrel{\spose{\lower 3pt\hbox{$\mathchar"218$}}
     \raise 2.0pt\hbox{$\propto$}}}

\def\F{{\bf F}}

\def\I{{\bf I}}

\def\M{{\bf M}}

\def\y{{\bf y}}

\def\l{\ell}

\def\c{{\bf c}}
\def\ch{\hat{\bf c}}

\def\nothing{\noindent\centerline{\,}}

\def\figone{
\begin{figure}
\makebox{
\noindent
\parbox[l]{3.0truecm}{\footnotesize
\caption{
The $\l_*=10$ window functions for the three methods discussed,
corresponding to using the original $\y$, Cholesky decomposing the Fisher 
matrix and taking its square root, respectively.
}
}
\hglue0.3cm
\parbox[r]{8.4truecm}{
\epsfxsize=8.4truecm\epsfbox{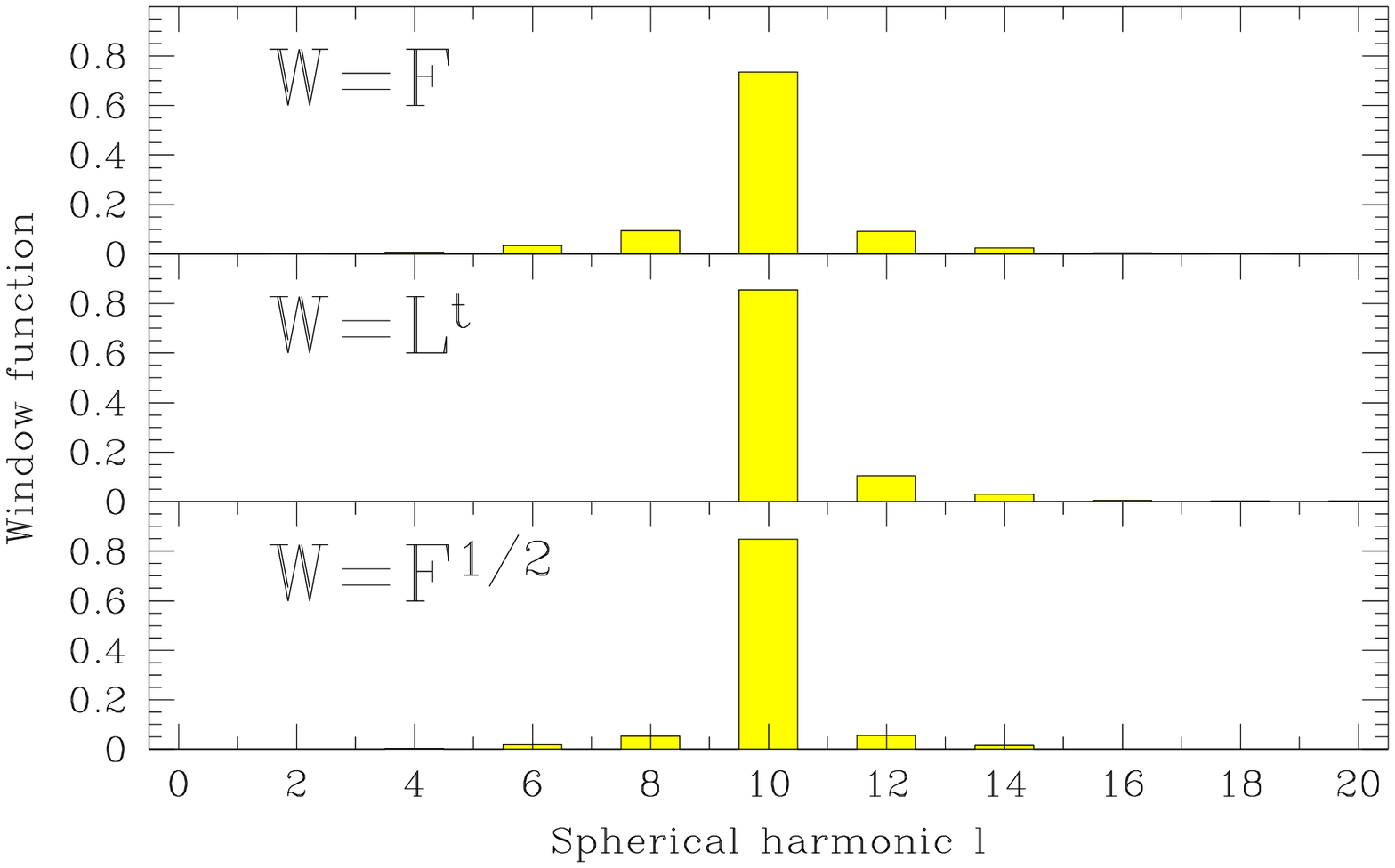}
\vskip-0.3truecm
}
}
\label{WindowFig}
\vskip0.3truecm
\end{figure}
}

\def\figtwo{
\begin{figure}
\makebox{
\noindent
\parbox[l]{6.0truecm}{\footnotesize
\flushleft
\epsfxsize=6.0truecm\epsfbox{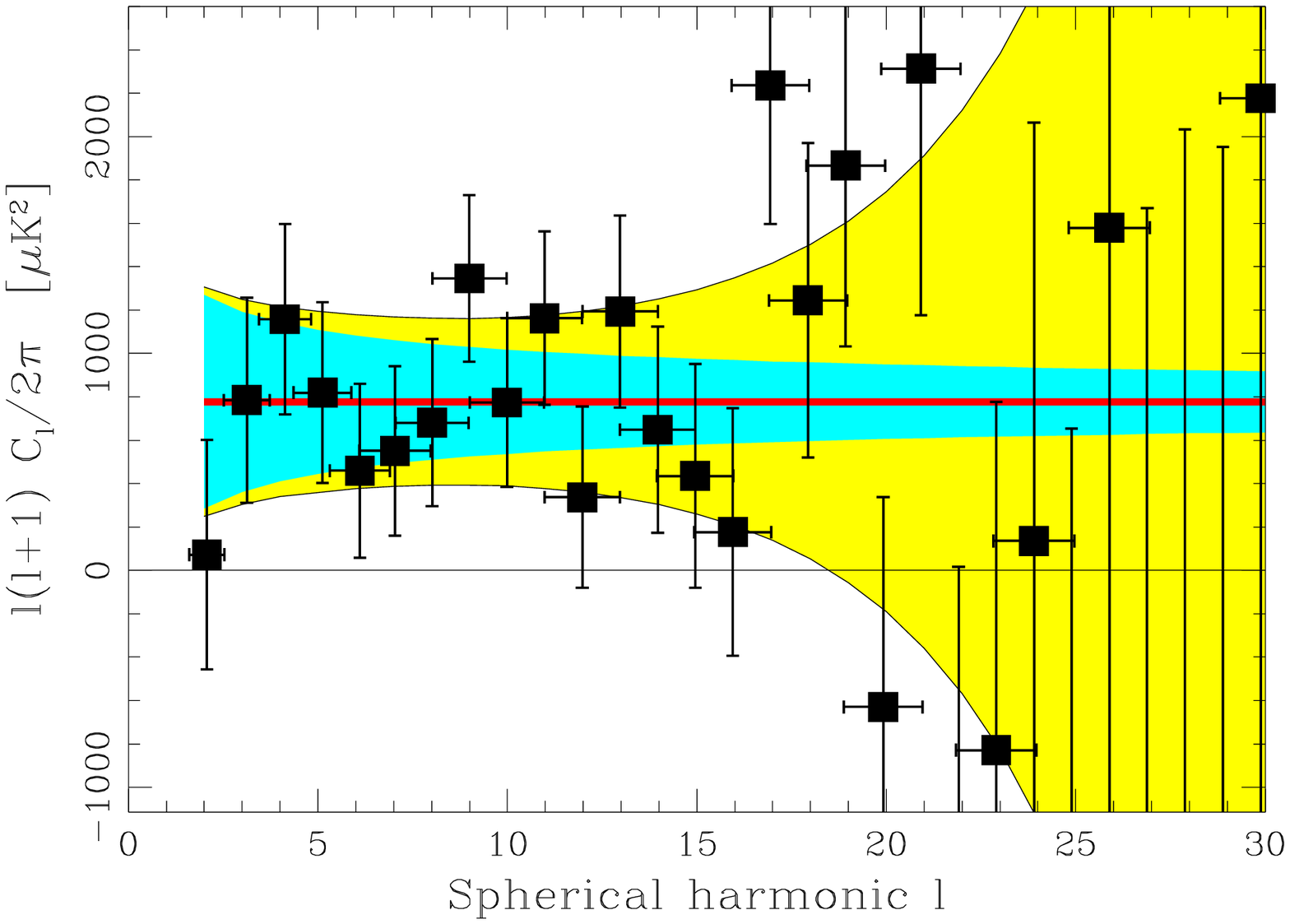}
}
\parbox[r]{5.7truecm}{
\epsfxsize=5.7truecm\epsfbox{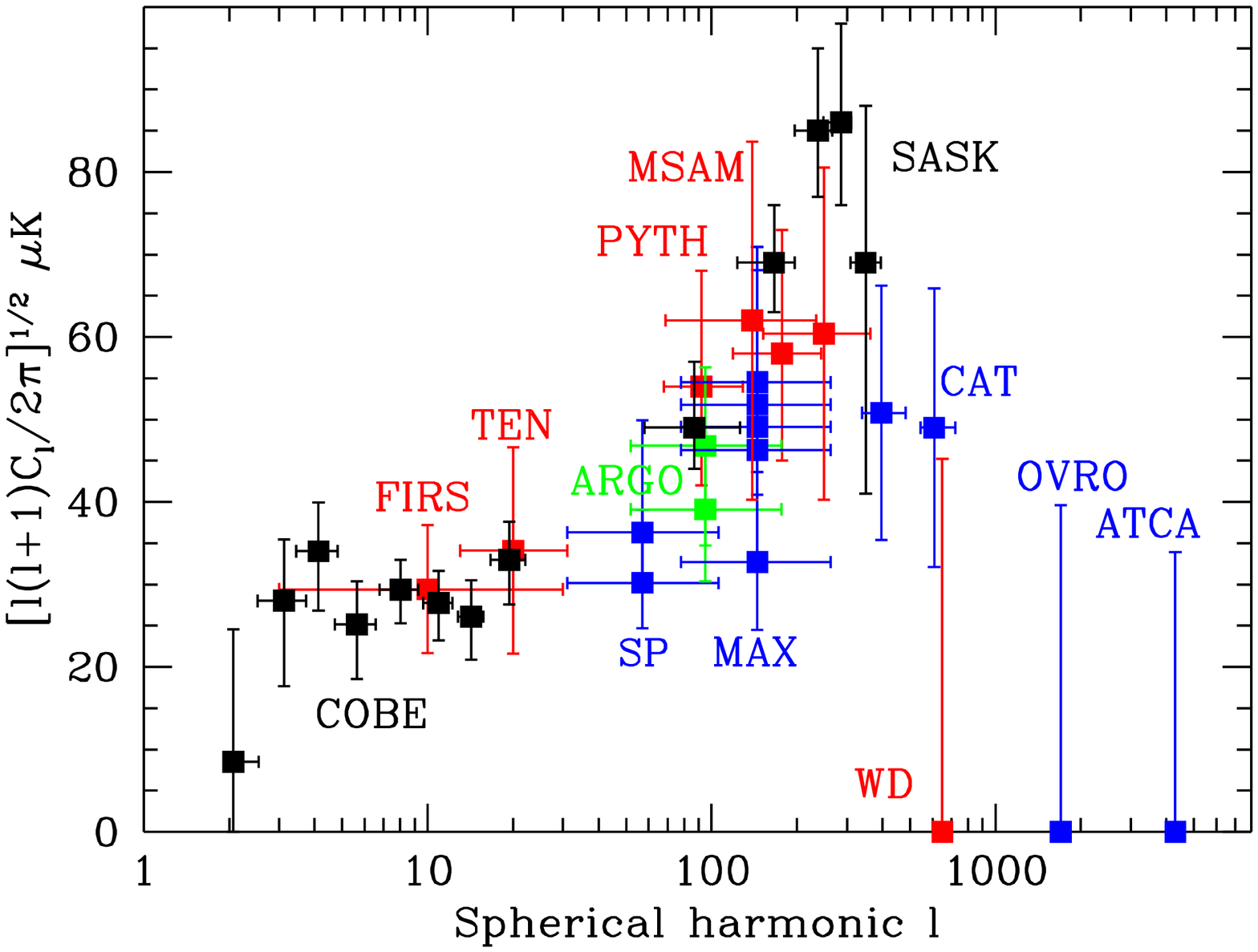}
}
}\\
\vskip-0.3truecm
\caption{
The power spectrum observed by COBE/DMR alone (left)
and binned into 8 bands and compared with other experiments (right). 
}
\vskip-0.3truecm
\label{ClFig}
\end{figure}
}

\documentstyle[sprocl,epsf]{article}

\begin{document}
\title{UNCORRELATED MEASUREMENTS OF THE CMB POWER SPECTRUM}
\author{ MAX TEGMARK~\footnote{Hubble Fellow} }
\address{Institute for Advanced Study, Princeton, NJ 08540, USA;
max@ias.edu}
\author{ A. J. S. HAMILTON }
\address{JILA and Dept. of Astrophysical, Planetary and Atmospheric Sciences, 
Box 440, Univ. of Colorado, Boulder, CO 80309, USA;
ajsh@dark.colorado.edu}
\maketitle
\abstracts{
We describe how to compute estimates of the power spectrum $C_\l$ from 
Cosmic Microwave Background (CMB) maps that not only retain all 
the cosmological information, but also 
have uncorrelated error bars and well-behaved window functions.
We apply this technique to the 4-year COBE/DMR data.
}

Accurate future measurements of the angular power spectrum $C_\l$ of the 
CMB would allow us to measure many key cosmological 
parameters with unprecedented accuracy~\cite{Jungman}. 
It has recently been shown~\cite{cl} how to compute a vector of
power spectrum estimates $\y$ retaining all the cosmological 
information from a CMB map, and whose mean and covariance is given by
\beqa{ExpecEq}
\expec{\y}&=&\F\c,\\\nonumber
\expec{\y\y^t}-\expec{\y}\expec{\y^t}&=&\F.
\eeqa
Here $\c$ is the vector of true power coefficients, {\ie}, 
$\c_\l=C_\l$, so the window function matrix and the 
covariance matrix are one and the same, equaling $\F$, 
the {\it Fisher information matrix}~\cite{cl}.
There are infinitely many ways of producing uncorrelated 
power estimates~\cite{H97b}. Making a factorization 
$\F=\M\M^t$ for some matrix $\M$, the new power estimates in the vector 
defined by
\beq{chDefEq}
\ch\equiv\M^{-1}\y
\eeq
will be uncorrelated, since $\expec{\ch\ch^t}-\expec{\ch}\expec{\ch^t}=\I$.
$\expec{\ch}=\M^t\c$, so the new window function matrix will be $\M^t$.
However, whereas the window functions of the original power estimates (the rows of $\F$)
are always well-behaved (they are always non-negative~\cite{cl}, and are generally
quite narrow, as shown in the top panel of Figure 1), 
there is no guarantee that the same will hold for the new
window functions. As was recently shown~\cite{H97b}, however, one generally
obtains beautiful window functions if one requires $\M$ to be lower-triangular, 
in which case $\F=\M\M^t$ corresponds to a Cholesky decomposition.
For the COBE/DMR case with a ``custom"~\cite{Bennett} 
Galaxy cut, this gives the narrow and non-negative window functions
in the middle panel of Figure 1, with side lobes only to the right.
Similarly, one could obtain window functions with
side lobes only to the left by chosing $\M$ upper-triangular.
A third choice, which is the one we recommend, is choosing 
$\M$ {\it symmetric}, which we write as $\M=\F^{1/2}$.
The square root of the Fisher matrix is seen to give 
beautifully symmetric window functions
(Figure 1, bottom) that are not only non-negative, but
also even narrower than the original (top), which has roughly the 
bottom profile convolved with itself.

\noindent\figone

\noindent\figtwo

\begin{table}
\caption{The COBE power spectrum 
$\delta T\equiv [\l(\l+1)C_\l/2\pi]^{1/2}$ in $\mu K$.}
$$
\begin{tabular}{|c|cccccc|}
\hline
Band&$\l_*$&$\expec{\l}$&$\Delta\l$&$\delta T$&$-1\sigma$&$+1\sigma$\\
\hline
1&2&2.1&0.5& 8.5& 0&24.5\\
2&3&3.1&0.6&28.0&17.7&35.5\\
3&4&4.1&0.7&34.0&26.8&40.0\\
4&5-6&5.6&0.9&25.1&18.5&30.4\\
5&7-9&8.0&1.3&29.4&25.3&33.0\\
6&10-12&10.9&1.3&27.7&23.2&31.6\\
7&13-16&14.3&2.5&26.1&20.9&30.5\\
8&17-30&19.4&2.8&33.0&27.6&37.6\\
\hline
\end{tabular}
$$
\label{Table1}
\nothing\vskip-1.0truecm
\end{table}

\nothing\vskip-1.0truecm
Figure 2 (left) shows the power spectrum extracted from the 4 year COBE
data~\cite{Bennett} with the minimum-variance method~\cite{cl} and 
de-correlated with $\M=\F^{1/2}$. 
The error bars are for a flat 18 $\mu$K spectrum.
These 29 data points thus contain all the cosmological information
from COBE, distilled into 29
mutually exclusive (uncorrelated) and collectively exhaustive (jointly 
retaining all information about cosmological parameters) chunks.
To reduce scatter, these have been binned into 8 bands in Table 1
and Figure 2 (right). The data and references for the other experiments 
plotted can be found in recent compilations~\cite{Lineweaver,Rocha}.

This method can readily be applied to other CMB experiments~\cite{Knox} as
well as galaxy surveys~\cite{H97b,H97a}. In comparison, 
previous CMB power spectrum 
estimation methods~\cite{Hinshaw,Wright,cobepow,Gorski} all had the drawback of giving  
correlated errors.

\smallskip
Support for this work was provided by
NASA through a Hubble Fellowship,
{\#}HF-01084.01-96A, awarded by the Space Telescope Science
Institute, which is operated by AURA, Inc. under NASA
contract NAS5-26555.
The COBE data sets were developed by the NASA
Goddard Space Flight Center under the guidance of the COBE Science Working
Group and were provided by the NSSDC.

\def\rf#1;#2;#3;#4;#5 {#1, {\it #3} {\bf #4}, #5 (#2).}

\end{document}